\documentclass[10pt,conference]{IEEEtran}
\IEEEoverridecommandlockouts
\usepackage{amsmath}
\usepackage{bm}
\AtBeginDocument{%
  \providecommand\BibTeX{{%
    Bib\TeX}}}
\usepackage{cite}
\usepackage{amsmath,amssymb,amsfonts}
\usepackage{algorithmic}
\usepackage{graphicx}
\usepackage{textcomp}
\usepackage{xcolor}
\usepackage{tabularx}
\def\BibTeX{{\rm B\kern-.05em{\sc i\kern-.025em b}\kern-.08em
    T\kern-.1667em\lower.7ex\hbox{E}\kern-.125emX}}
\usepackage{tcolorbox}
\usepackage{multirow}
\usepackage[normalem]{ulem}
\useunder{\uline}{\ul}{}

\begin{document}
\author{\IEEEauthorblockN{Nusrat Zahan,
Shohanuzzaman Shohan, Dan Harris and
Laurie Williams}
\IEEEauthorblockA{North Carolina State University\\
Raleigh, USA\\
Email: [nzahan, sshohan, doharris, lawilli3]@ncsu.edu
}}
\newcommand{\realsearchgoal}{\textbf{\textit{ The goal of this study is to assist practitioners and researchers in making informed decisions on which security practices to adopt through the development of models between software security practice scores and security vulnerability counts.
}}}

\newcommand{\rqPattern} {Do packages with higher aggregate security scores have fewer vulnerabilities?}
\newcommand{\rqOutcome}{Which Scorecard security practices are most important to understand the relationship between security practices and vulnerability counts in regression models?}

\title{Do Software Security Practices Yield Fewer Vulnerabilities?} 
\maketitle
\begin{abstract}
Due to the ever-increasing number of security breaches, practitioners are motivated to produce more secure software. In the United States, the White House Office released a memorandum on Executive Order~(EO) 14028 that mandates organizations provide self-attestation of the use of secure software development practices.  The OpenSSF Scorecard project allows practitioners to measure the use of software security practices automatically. %
However, little research has been done to determine whether the use of security practices improves package security, particularly which security practices have the biggest impact on security outcomes. \realsearchgoal 

To that end, we developed five supervised machine learning models for npm and PyPI packages using the OpenSSF Scorecard security practices scores and aggregate security scores as predictors and the number of externally-reported vulnerabilities as a target variable. %
Our models found that four security practices (\textit{Maintained, Code Review, Branch Protection, and Security Policy)} were the most important practices influencing vulnerability count. However, we had low $R^2$ (ranging from 9\% to 12\%) when we tested the models to predict vulnerability counts. Additionally, we observed that the number of reported vulnerabilities increased rather than reduced as the aggregate security score of the packages increased. Both findings indicate that additional factors may influence the package vulnerability count. %
Other factors, such as the scarcity of vulnerability data, time to implicate security practices vs. time to detect vulnerabilities, and the need for more adequate scripts to detect security practices, may impede the data-driven studies %
to indicate that a practice can aid in the reduction of externally-reported vulnerabilities. We suggest that vulnerability count and security score data be refined such that these measures may be used 
to provide actionable guidance on security practices. %
\end{abstract} 

\section{Introduction}
A 2022 report from Synopsys~\cite{Synopsys_report} assessed the reliance of the software industry on open-source software (OSS) and estimated that 97\% of applications use OSS. %
The continuous reliance on OSS comes with a risk of supply chain attack. %
Sonatype has recorded an average 700\% jump in supply chain attacks~\cite{sonatype_2022}, as measured by the number of newly-published malicious packages in open-source repositories. Software developers largely did not anticipate how the software supply chain would become a deliberate attack vector. %
Many stakeholders have started to recognize this urgent concern; most prominently, Executive Order 14028~\cite{EO_2021}, issued May 12, 2021, calls for identifying practices that enhance software security and mandating the use of software security practices.%

As organizations seek to address escalating security risks and comply with regulations, a myriad of activities are available to improve software security. The National Institute of Standards and Technology (NIST) issued guidance~\cite{souppaya2022secure,NIST_2} on software development practices that enhance the security of the software supply chain. %
The Open Source Security Foundation (OpenSSF), a cross-industry organization hosted at the Linux Foundation, provides the vehicle for collaboration on tools, services, training, infrastructure, and resources for securing open-source projects. The OpenSSF Scorecard~\cite{Scorecard} project computes automated scores of package's security practices to help developers make better decisions about security when consuming open source projects.   %

While NIST's guidelines and OpenSSF projects provide guidelines and comprehensive lists of security practices, challenges arise in validating whether these practices improve security. Organizations may not have the resources to adopt a full suite of these practices. They would like to understand the key drivers of success, especially which of many possible software security practices to undertake first.  A data-driven study on the relationship between the use of software security practices and vulnerability count could aid in this understanding. \realsearchgoal %

To that end, this study is focused on the relationship between publicly-available data of security practice use and externally-reported vulnerability count. %
This relationship may be used to provide actionable recommendations to practitioners on security practices. Our work addresses the following \textbf{research questions}:

\begin{itemize}
    \item \textbf{RQ1 (Security Practices)}: \rqOutcome \label{RQ2} 
    \item \textbf{RQ2 (Security Outcome)}: \rqPattern \label{RQ1}
\end{itemize}

The OpenSSF Scorecard project~\cite{Scorecard} uses security practice metrics and auto-generates a ``security score'' for each practice. The tool also computes an aggregate ``security score'' which is a weighted average of the individual score. %
In this study, we leverage the security practice score data provided by the OpenSSF Scorecard project and vulnerability information from security advisory databases to evaluate the relationship between security practices and vulnerability count. To answer RQ1, we used the ``feature importance'' technique in four regression models to understand the importance of each security practice in interpreting the relationship between security practices and vulnerability count. For RQ2, we built our fifth regression model to evaluate the statistical association of a package's aggregate security scores with package vulnerability count and to understand whether implementing security practices assists in secure coding with fewer vulnerabilities.

This study can assist practitioners in making informed choices regarding their security practices, particularly on which practices have the most impact or whether specific security practices have improved security outcomes. This study makes the following contributions:
\begin{itemize}
    \item A proposed model to identify important security practices to enhance software security.
    \item A proposed model to understand how higher aggregate security scores affect security outcomes.
    \item Evaluation of the proposed model and security practices using open source datasets. 
    \item The list of challenges impeding the model's performance
\end{itemize}

This paper is organized as follows: Section \ref{background} includes the related concept used in our study,  section \ref{related_work} discussed the step taken by the software industry to improve software supply chain security. Section \ref{Data} describes the data collection process, and Section \ref{experiment} describes the data analysis to measure model performance. We close with a discussion and limitations of our findings (Section \ref{Discussion}).

\section{Background} \label{background}
In this section, we discuss the ``OpenSSF Scorecard'' tool and the two most frequently used terms in our study: ``Security Outcome'' and ``Package''.
\subsection{\textbf{OpenSSF Scorecard}}\label{Scorecard_metrics}
The OpenSSF Scorecard~\cite{Scorecard} is an automated tool that runs on source code hosted by GitHub to monitor the security health of the packages. The Scorecard automatically computes a normalized integer score between 0 to 10 for each of the 18 security practices. %
For each GitHub repository, an individual score for each practice and an aggregate score for all practices are returned by the tool. Each security practice is assigned one of four risk levels: \textbf{`` Critical"} risk-weight 10; \textbf{`` High”} risk-weight 7.5; \textbf{`` Medium”} risk-weight 5; and \textbf{`` Low”} risk-weight 2.5. The aggregate score is a weight-based average of the individual scores, weighted by risk.  

Apart from the scores between 0-10, the Scorecard tool also assigns a score of -1 to the project. The notation -1 indicates that Scorecard could not get conclusive evidence of implementing practices, or perhaps an internal error occurred due to a runtime error in Scorecard. The inconclusive outcome is graded as -1 instead of 0. Since a value of 0 will affect the package’s aggregate score, Scorecard assigned a value of -1 to avoid the penalty of failing a check.

\subsection{\textbf{Security Outcome}}
The Security Outcomes represent indications of the failure or achievement of security associated with the software throughout the software’s life cycle. Counting vulnerabilities is the most common means of measuring security in software~\cite{morrison2014mapping,sec_outcome,morrison2015security}. %
In this study, we used package vulnerability counts as a security outcome. 

\subsection{\textbf{Package}}
A \textbf{Package Registry} such as npm and PyPI stores packages, the metadata associated with them, and the configurations needed to install it, as well as keeps track of the versions of packages~\cite{package_2}. The registry offers a way to upload packages and APIs for the client to call packages. Anyone can download and reuse the packages from the registry. A \textbf{Package} is a reusable piece of software that can be called from a global registry (npm, PyPI) as a dependency to build another package~\cite{package_1,package_2}. Each package may or may not depend on other packages. However, publishing a package in a registry allows consumers to import and reuse these packages.

\section{Related Work} \label{related_work}
This section highlights prior related studies, different security guidelines, frameworks, and tools on package security health and security practices. While we only studied the Scorecard tool's security metrics, we further discussed other security frameworks and tools to encourage future research to evaluate which of these recommended security practices help to improve security outcomes. Because many practices are recommended for software security, but none of these guidelines (\ref{guidelines}) demonstrated which of these practices improved security and why we should adopt them. Practitioners will comply with these guidelines if data-driven research studies on security practice measurement and feedback loops validate the improvement in security outcomes. %

\subsection{\textbf{Research Studies}} \label{Scorecard_related}

Zahan et al.~\cite{Zahan2022} conducted a measurement study on the \textbf{OpenSSF Scorecard} security practices in the context of the npm and PyPI ecosystems to identify the ecosystem's security practices trends and gaps. The study analyzed Scorecard scores for 767,389 npm packages and 191,158 PyPI packages. The authors observed gaps in both ecosystems in practicing \textbf{Code-Review, Maintained, License, Branch-Protection}, and \textbf{Security-Policy} practices in the GitHub repository. Then, the rules specified by the Scorecard tool for Dependency-Update-Tool, Fuzzing, CII-Best-Practices, Signed-Releases, and Packaging security checks, were weakly adopted in npm and PyPI due to Scorecard's inherent reliance on other systems. The study revealed that 13 Scorecard metrics could be mapped back to the NIST SSDF framework's~\cite{souppaya2022secure} security practices. Although this research showed a gap in both ecosystems to adopt different security practices, the authors did not verify whether adopting these practices would improve the npm and PyPI ecosystems' security outcomes. %

On October 18, 2022, Sonatype released the \textbf{8th annual State of the Software Supply Chain report}~\cite{sonatype_8}. As part of project quality metrics identification, they built several machine learning classification models on Scorecard security practices to understand how well a model based on security practices could correctly identify projects with known vulnerabilities. Their classification models were based on 12,786 Java projects, and they found \textbf{Code-Review, Binary-Artifacts, Pinned Dependencies, Branch Protection} as important practices in their classification model.

Our study focuses on understanding whether good security practices improve security outcomes, whereas Sonatype research built models to identify vulnerable projects using Scorecard data. However, study~\cite{Zahan2022} showed that not all Scorecard metrics apply to npm and PyPI ecosystems. From the Sonatype annual report, we could not verify whether all the metrics applied to Java projects or whether the study removed noisy metrics from the model's independent variables.

\subsection{\textbf{Guidelines and Frameworks: }} \label{guidelines}
The \textbf{Software Component Verification Standard (SCVS)}~\cite{scvs} by OWASP, is a framework to develop a common set of activities, controls, and security practices that can help in identifying and reducing risk in a software supply chain. There are $6$ control families that contain $87$ controls for different aspects of security verification or processes. The SCVS has three verification levels, where higher levels include additional controls. 

The \textbf{Building Security In Maturity Model (BSIMM)}\cite{BSIMM} is the result of a multiyear study of real-world software security initiatives (SSIs). Each year, various firms in different industry verticals use the BSIMM to manage their SSI improvements because the BSIMM report provides a clear picture of actual practices used by organizations across the security landscape~\cite{weir2021infiltrating}. The BSIMM is organized as a set of 122 activities, which consist of four domains distributed into 12 practices.

The \textbf{CNCF Technical Advisory Group (TAG)}~\cite{CNCF_TAG} published a series of recommended best practices, tooling recommendations, and design considerations that can reduce the likelihood and overall impact of a successful supply chain attack. They discuss the $5$ stage of software supply chain security—securing code, materials, building pipelines, artifacts, and deployments. The framework proposed $54$ practices with associated risk factors to provide a holistic, end-to-end guide for organizations and teams.

In response to Section 4 of the President’s Executive Order (EO) on “Improving the Nation’s Cybersecurity (14028)”~\cite{EO_2021}, the U.S. National Institute of Standards and Technology (NIST) improvised the \textbf{Secure Software Development Framework (SSDF)}~\cite{souppaya2022secure}. The framework does not introduce new practices or define new terminology. Instead, the framework describes a set of high-level practices based on established standards~\cite{BSIMM,BSA,OWASP_samm,CNCF_TAG} of secure software development practices. %

\begin{table*} \renewcommand{\arraystretch}{1.2} 
\caption{\centering{Scorecard Security Practices used in ~\cite{Zahan2022} }} \label{tab:scorecards} 
\centering
\begin{tabular}{| p{10pt} || p{95pt} || p{280pt} ||p{40pt}| }\hline
No. & Security Practices & Description & Risk Label \\
\hline\hline
1& Dangerous-Workflow & This check determines dangerous code patterns in the project's GitHub Action workflows. These patterns include logging GitHub context and secrets, untrusted code checkouts, or script injection with untrusted context variables.  & Critical\\\hline

2& Branch-Protection & This check determines whether a project's branches are protected with GitHub's branch protection settings. This check has five tiers of scoring. Each tier must be fully satisfied to acquire points at the next tier.  & High\\\hline

3& Code-Review & This check determines whether the project practices code review before merging pull requests. The check first detects whether Branch-Protection is enabled with at least one required reviewer. If this fails, the check determines whether the most recent commits are Github-approved reviews or whether the merger differs from the committer. Note that this check is not applicable if the package has one maintainer. & High\\\hline

4& Binary-Artifacts &This check determines whether the project has generated executable (binary) artifacts in the source repository since Binary-Artifacts cannot be reviewed, allowing possible obsolete or maliciously subverted executables. &High\\\hline

5& Dependency-Update-Tool & This check determines whether the dependency updates tools like dependabot or renovatebot are enabled or not. A project that uses other tools will receive a low score on this test.  & High\\\hline

6& Maintained & This check determines whether the project is actively maintained. If there is activity on commit and issues from users who are collaborators, members, or owners, the project receives the highest and partial score. & High\\\hline

7& Signed-Releases & This check determines whether the project signs release artifacts, currently limited to GitHub repositories only. This check looks for the following filenames in the project's last five releases: *.minisig, *.asc (pgp), *.sig, *.sign. & High\\\hline

8& Token-Permissions & This check determines whether the project's automated workflow tokens are set to read-only. If the permissions definitions in each workflow's yaml file are set as read-only at the top level, and the required write permissions are declared at the run-level, the project gets the highest score.  If the top-level permissions are not defined, one point is reduced.  &High\\\hline

9& Vulnerabilities & This check determines whether the project has unfixed vulnerabilities using the OSV (Open Source Vulnerabilities) database.  &High\\\hline

10& Pinned-Dependencies &This check determines if the project has pinned its dependencies. The check works by looking for unpinned dependencies in Dockerfiles, shell scripts, and GitHub workflows. & Medium\\\hline

11& Packaging & This check determines if the project is published as a package. The check looks for GitHub packaging workflows and language-specific GitHub Actions that upload the package to a corresponding hub. & Medium\\\hline

12& Fuzzing & This check determines if the project uses fuzzing by checking the repository name in the OSS-Fuzz project list. & Medium\\\hline

13& Security-Policy & This check determines if the project has published a security policy  by looking for a file named SECURITY.md. & Medium\\\hline

14& License & This check determines if the project has published a license by checking standard locations for a file named according to common license conventions. & Low \\\hline

15& CII-Best-Practices & This check determines whether the project has a CII Best Practices Badge to ensure that the project uses a set of security-focused best practices. & Low \\\hline
\end{tabular}
\end{table*}
\textbf{Supply Chain Levels for Software Artifacts (SLSA)}~\cite{SLSA} is a security framework established by industry consensus, a set of standards and controls to prevent tampering, enhance the integrity, and also a set of security rules that may be adopted incrementally. SLSA has four levels, with SLSA 4 representing the ideal end state. The lower levels represent incremental milestones with corresponding incremental integrity guarantees. 

In August 2022, the Cybersecurity and Infrastructure Security Agency (CISA) and the National Security Agency (NSA) released a recommended practice guide on \textbf{Securing the Software Supply Chain for Developers}~\cite{CISA_dev}. The report advocated specific frameworks like SLSA and SSDF. The framework is a roadmap to building a healthy software development environment that includes a Software Bill of Materials (SBOM) for describing software components, active monitoring for vulnerabilities and software supply chain attacks, and a secure build environment and development team. 

Microsoft released the \textbf{Open Source Software (OSS) Secure Supply Chain (SSC) Framework}~\cite{Microsoft_framework}, which is a security assurance and risk reduction process focused on securing how developers consume open source software. The framework provides security guidance and tools throughout the developer's inner-loop and outer-loop processes that have played a critical role in defending and preventing supply chain attacks through the consumption of open-source software across Microsoft. In September 2022, the OpenSSF released the \textbf{npm Best Practices Guide}\cite{npm} to help JavaScript and TypeScript developers reduce the security risks associated with using open-source dependencies.

\subsection{\textbf{Tools}} \label{tool}

\textbf{Open Source Insights (OSI)} or Deps.dev~\cite{OSI} is a Google-developed and hosted tool to aid practitioners in grabbing information about the source code location, package metadata, licenses, releases, and vulnerabilities of open-source products. OSI scans millions of open-source packages from different ecosystems, constructs dependency graphs, and annotates the metadata in a dashboard. Apart from the package's metadata, the OSI dashboard also shows statistics about a package's direct or transitive dependencies. On the OSI website, users can view the vulnerability mapping of a package as well as the vulnerability mapping with associated dependencies. In addition to the package metadata, OSI has also incorporated Scorecard security practices to help understand package security practices.

\textbf{In-toto:} A framework that holistically enforces the integrity of a software supply chain by gathering cryptographically verifiable information about the chain itself~\cite{torres2019toto,in-toto}. \texttt{in-toto} grants the end user the ability to verify the software’s supply chain from the project’s inception to its deployment. To achieve this, \texttt{in-toto} requires a project owner to declare and sign a layout of how the supply chain’s steps must be carried out and by whom. The concerned parties will document their actions and produce a cryptographically signed declaration for each step as they are completed. The link metadata recorded from each step can be verified to ensure that each step was completed correctly. Both SLSA~\cite{SLSA} and CNCF Technical Advisory Group (TAG)~\cite{CNCF_TAG} recommend in-toto as a framework to secure the integrity of software supply chains.

\begin{figure*}[!htb]  
   \centering
   \includegraphics[width=6.0in]{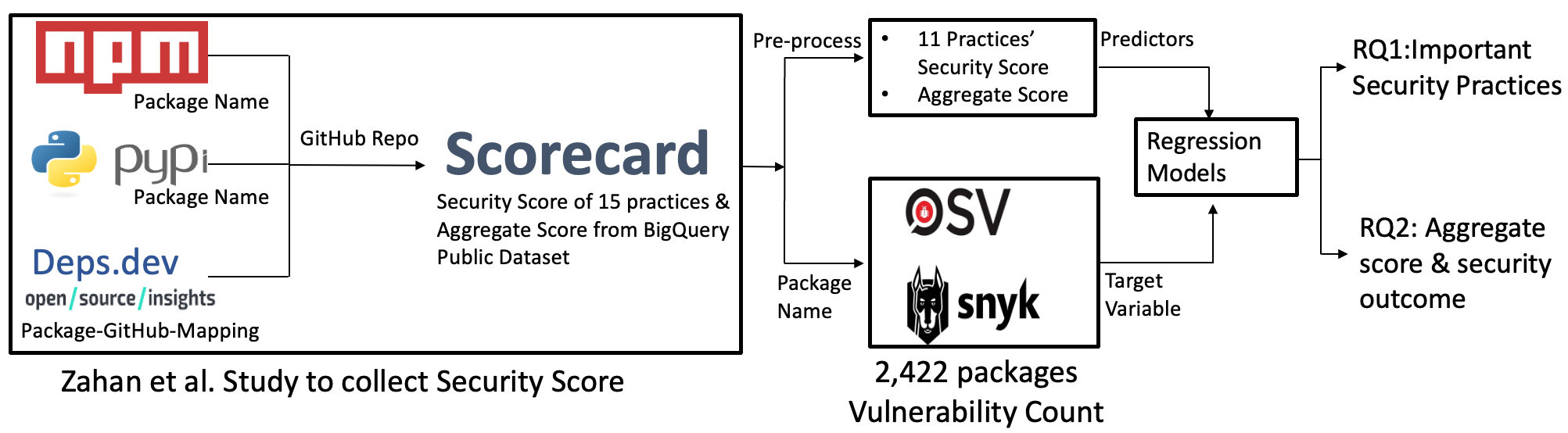} 
\caption{Research Workflow}
\label{fig:method}
\end{figure*}

\section{Study Design and Data Collection} \label{Data}

In this section, we discuss our study design and the data collection process that was used in our research. %
 Section \ref{design} highlights our high-level overview of the research workflow. Section \ref{Scorecard_metrics} covers how we obtained Scorecard security practices' scores for each package, and Section \ref{vulnerbilities} discusses how we collected vulnerability count data from Open Source Vulnerabilities (OSV) and Snyk databases.  

 \subsection{\textbf{Study Design}} \label{design}
To answer our two research questions, we leverage regression models where we used the security scores of package repositories' as independent variables and security outcomes in terms of package vulnerabilities count as the dependent variable. Figure \ref{fig:method} shows an overview of the research workflow.

RQ1 aims to identify the most important practices that explain the relationship between security practices and security outcomes. To that, we utilized the 15 security practice score as independent variables generated by the Scorecard tool. Next, our RQ2 verifies whether packages with higher aggregate security scores have fewer vulnerabilities. Here, we leverage each repository's aggregate security score as an independent variable, which is generated by Scorecard by measuring the weight-based average of
the individual security practice's scores, weighted by risk. For both RQ1 and RQ2, we used the security outcomes in terms of vulnerability counts as the dependent variable. 

For RQ1, we classified a security practice as an important metric if it is identified as important in more than one model. To that, we analyzed 4 different regression model responses. For RQ2, we developed our fifth regression model to test our hypothesis.

\subsection{\textbf{Scorecard Security Practices data}} We obtained Scorecard security practices data from ~\cite{Zahan2022}. %
(see Section \ref{Scorecard_related}). The data includes the 15 security practices scores and aggregate security score of 767,389 npm packages and 191,158 PyPI packages. The dataset contained mapping between the package name, the GitHub repository, the 15 security practices, and the aggregate security score. Each practice score ranges between $-1$ to $10$. We describe the OpenSSF Scorecard project in Sections \ref{Scorecard_metrics}. %
The 15 security practices used in our study are discussed in Table \ref{tab:scorecards}. More details of these checks can be found in the Scorecard Project GitHub repository~\cite{Scorecard}. 

 The Scorecard tool has 18 security practices. However, the OpenSSF Scorecard team publishes the security score data for 15 of these practices. The Scorecard team took out CI-Test, SAST, and Contributor practices to scale the weekly job since computing these practices is API intensive.%
 ~The security scores for the other three practices can be obtained by running Scorecard from the command line interface (CLI). Our study can be replicated for 15 security practices by collecting weekly data from the Google BigQuery dataset published by the Scorecard team. 

\begin{table*}[!t] \centering  \renewcommand{\arraystretch}{1.2} \caption{Unique vulnerabilities count collected from OSV and Snyk} \label{tab:vul_details}
\footnotesize
\begin{tabular}{|c||c||cc||cc||cc||c|c|} \hline 
Ecosystem & Database  & Github\_Package & GitHub\_Vul\_Count & CVE & Non-CVE & Total\_GitHub\_Package & Total\_GitHub\_Vul \\\hline\hline
\multirow{2}{*}{PyPI} & Snyk  & 523 & 1189 & 830 & 359 & \multirow{2}{*}{589} & \multirow{2}{*}{2261} \\
 & OSV  & 352 & 1407 & 1374 & 33 &  &  \\\hline
\multirow{2}{*}{npm} & Snyk  & 1689 & 2838 & 1888 & 950 & \multirow{2}{*}{1833} & \multirow{2}{*}{3655} \\
 & OSV  & 1311 & 1883 & 1559 & 324 &  & \\\hline
\end{tabular}
\end{table*}

\subsection{\textbf{Advisory Data}} \label{vulnerbilities}
This section provides background on the database used in this study and advisory data collection processes to obtain the vulnerability count.

\subsubsection{\textbf{Database}} \label{OSV_snyk}
We used the OpenSSF Open Source Vulnerabilities (OSV)~\cite{osv_adv} and Snyk.io vulnerability~\cite{Snyk_adv} databases in this study. Both databases have a similar structure in terms of the patch, Common Vulnerabilities and Exposures (CVE), and Common Weakness Enumeration (CWE) information and they track vulnerabilities across multiple ecosystem's databases. In September 2022, we acquired vulnerability data from these two databases for both ecosystems. 

\textbf{OSV} provides both human and machine-readable data formats to describe vulnerabilities and maps to open-source package versions or commit hashes.  OSV is an aggregator of vulnerabilities from different security advisories, e.g., GitHub Security Advisories~\cite{GHSA}, PyPI Advisory Database~\cite{pypi_adv}, Go Vulnerability Database~\cite{go_adv}, Rust Advisory Database~\cite{rust_adv}, OSS-Fuzz (mostly C/C++)~\cite{ossfuzz_adv}, and Global Security Database~\cite{gsd_adv}. These databases have adopted the OpenSSF OSV format to facilitate use in tooling and aggregation. \textbf{Snyk.io} is another platform that provides security reports containing vulnerability information for different package ecosystems, including npm, PyPI. %

\subsubsection{\textbf{Vulnerability Data Collection}}
As Scorecard only works for Github repositories, our vulnerability collection was limited to packages with GitHub repositories. Hence, each database may have more vulnerabilities than we presented in this study. Table \ref{tab:vul_details} contains the detailed counts of vulnerabilities. 

The~\cite{Zahan2022} data contained 767,389 npm and 191,158 packages with GitHub repositories. We extract the vulnerability data from both databases using the package name. For each package, we collected vulnerability details in terms of CVE, CWE and patch link from both databases. OSV exported data to a Google Cloud Storage (GCS) bucket. The bucket contains a zip of all vulnerabilities JSON file for each ecosystem at \texttt{gs://osv-vulnerabilities/<ECOSYSTEM>/all.zip}. However, we had to build a scraper for Snyk to extract data from the Snyk website.

To create the dataset, we counted the number of unique vulnerabilities found for each package. First, we merged both databases and applied our filter conditions.  During this stage, we eliminated vulnerabilities if we found i) duplicate package-to-CVE mappings or duplicate patches for the Non-CVE vulnerabilities; ii) Non-CVE vulnerabilities without patch links; iii) Vulnerabilities with Non-CVE, NON-CWE and No Patch link. Table \ref{tab:vul_details} shows the unique count for each row. After filtering and merging both advisory databases, we found 589 PyPI packages with 2261 vulnerabilities and 1,833 npm packages with 3,655 vulnerabilities. In total, we have vulnerability counts of 2,422 packages as a dependent variable. 

At this stage, we mapped 2,422 packages with the Zahan et al.~\cite{Zahan2022} dataset to collect the Scorecard security practice's values. As a result, 2,422 packages with vulnerability count and security score are used as our final dataset, and we did not consider any other packages from~\cite{Zahan2022} study. 

\subsection{\textbf{Dependents and Downloads Data}} The number of dependents and downloads reflects the importance of a project based on how many other projects use it. To determine whether the packages in our dataset are utilized by other open-source packages, we obtained dependents and downloads data. We collected direct-dependent and transitive-dependent information from the Open Source Insights (OSI) API (Section \ref{tool}). %
Then, we collected 12 months of download statistics of npm packages from the study~\cite{zahan2022weak} and for PyPI packages from the Google BigQuery public dataset.%

\section{Data Analysis} \label{experiment}
This section discusses our data analysis to answer our research questions. %
Sections \ref{RQ1} and \ref{RQ2_ans} discuss pre-processing and regression models to answer our research question. %

\subsection{\textbf{RQ1: Important Security Practices}} \label{RQ1}
In this section, we discuss the analysis approach for RQ1: \textit{\rqOutcome}

\subsubsection{\textbf{Data Pre-processing}}\label{Preprocess}
In this section, we discuss the data pre-processing of missing, inconsistent, and noisy data found in Scorecard data due to internal and external errors. For example, we eliminated predictors and packages from our dataset if Scorecard generated a score of $-1$.  The score $-1$ depicts that the tool could not obtain concrete evidence that practice had been implemented or if an internal error had occurred due to a runtime error.  %
Therefore, we utilized the following exclusion criteria for security practices and packages to acquire consistent security scores. 
\begin{itemize}
    \item \textbf{Practice Exclusion}: We removed security practices if a practice has more than 10\% packages with $-1$ score. Signed-releases~(91\%), Packaging~(94\%) and Fuzzing~(16\%) practices had more than 10\% packages with a score of $-1$. We also removed ``Vulnerability'' metrics as a predictor. The metric checks whether a package has open or unfixed vulnerabilities in the OSV database. %
    Since part of our dependent variable was collected from OSV, we removed this metric to avoid introducing bias. Although we did not find any package in our dataset where a package had open vulnerabilities, it may occur in any future packages. To standardize our study for replication, we removed the vulnerability metric from our dataset. At this stage, we removed 4 practices and have 11 practices as predictors.
    
    \item \textbf{Package Exclusion}: We found Branch Protection~(6\%), and Code-Review~(0.08\%) had less than 10\% of packages with a score $-1$. Therefore, we only removed those packages instead of removing the practice itself. At this stage, we removed 150 packages from the dataset.
\end{itemize}

To protect the quality of the data that can directly affect how well our ML model learns, we are removing these practices. Given that Scorecard will compute the accurate security status of these practices in the future, practitioners can elect them when constructing our model. Conversely, leveraging different vulnerability databases other than OSV would allow us to use vulnerability practice as one of the predictors. However, the lack of vulnerability data compelled us to gather vulnerabilities from several sources.

Then, we used \texttt{qq-plot}~\cite{qq-plot} to determine outliers in vulnerability count data. Four packages had vulnerability counts (36, 77, 158, 217) far from the reference lines and were considered outliers. %
Since we will use regression models, extreme outliers will affect the line of regression. Hence, we removed those packages from our dataset as an outlier in the dependent variable. We utilized 2268 packages and 11 security practices in our model.

\begin{figure*}[h]  
   \centering
   \includegraphics[width=7.0in]{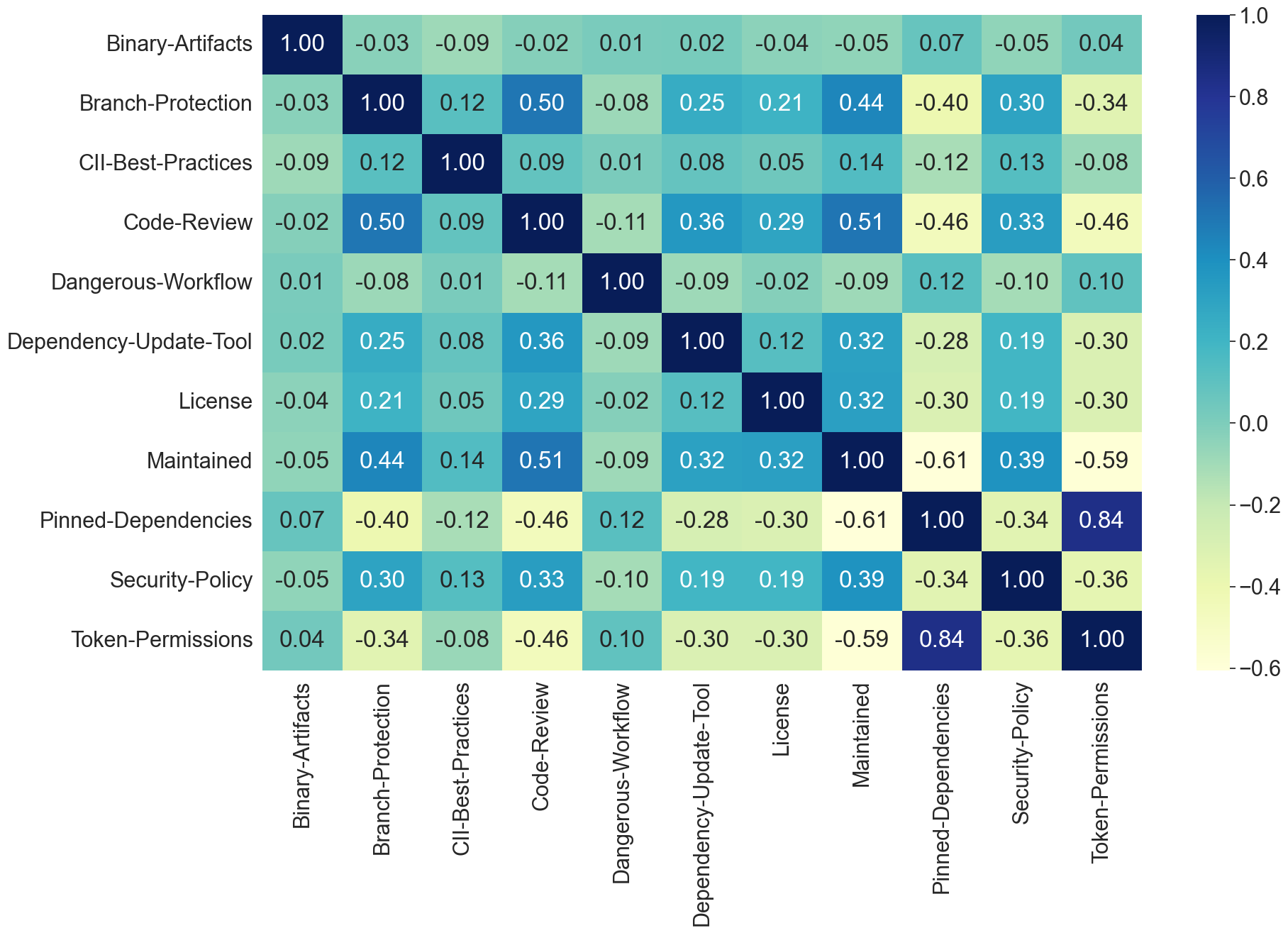} 
\caption{Heatmap matrix for the Spearman rank correlation test}
\label{fig:corr}
\end{figure*}
\subsubsection{\textbf{Correlation Analysis}} \label{cor}
To evaluate the normality of the dataset, the Henze-Zirkler multivariate normality~\cite{henze1990class} test was utilized. Results indicated that the dataset was not normally distributed. We expected non-normality, given that the independent variables were discrete variables. At this stage, we measure the correlation to know the association or the absence of a relationship between two variables in our dataset. With evidence of non-normality, we deployed the Spearman rank correlation~\cite{zar2005spearman} tests.%

A value close to +1 and -1 indicated a strong positive  or negative correlation, respectively. A correlation coefficient close to 0 indicates a weak correlation between the variables~\cite{ratner2009correlation}. Since values above $\pm 0.3$ start to indicate a moderate positive (negative) linear relationship~\cite{ratner2009correlation}, we consider values between $\pm 0.3$ to $\pm 1$ as a limit. Figure \ref{fig:corr} shows the heatmap matrix for the Spearman rank correlation test.

The Spearman test showed the strongest correlation between Pinned-Dependencies and Token-Permission(0.84), followed by Code-Review and Maintained (0.51), Branch-Protection showed a moderate correlation with Code-Review (0.50), and Maintained (0.44) and Security Policy (0.3). %
Negative correlations were observed between Maintained and Pinned-Dependencies (-0.61), Maintained and Token-Permission (-0.59), Code-Review and Pinned-Dependencies (-0.46), and Code-Review and Token-Permissions (-0.46). 

Token Permission looks at the permission of the GitHub workflow files, and Pinned Dependencies look for the presence of unpinned dependencies in those files, indicating why they might have a strong positive relationship. Both the Code-Review and Maintained variables showed a moderate positive relationship. In contrast, an increase in Maintained comes with a decrease in Pinned-Dependencies and Token-Permission practices. Token-Permission and Pinned-Dependencies do not require active maintenance directly, which might explain the inverse relationship. However, code review could be part of active maintenance, hence the positive relationship. Branch-Protection enables a formal approval process for code changes and pairs well with Maintained and Code-Review.

\subsubsection{\textbf{Feature Selection}}
For the selection of best features, we used the \texttt{SelectKBest}~\cite{selectK} library from the $sklearn$, which shows the feature importance based on $K$ highest scores. F-regression was selected as the scoring function, which performs a univariate linear regression test and returns F-statistics and p-values. All of the independent predictors were considered to identify the most important features (K=``all''). The SelectKBest analysis identified nine predictors as the most important features: Maintained, Security-Policy, Pinned-Dependencies, Code-Review, Branch Protection, Token-Permission,  License, CII-Best-Practices and Dependency-Update-Tool. Each of them had a high F-score and statistically significant p-values (p$<$0.01). 
A reduced dataset was prepared from the statistical significance measure with the selected components as independent variables and the total vulnerability count as the dependent variable. 
\begin{table*}[h] \centering \renewcommand{\arraystretch}{1.3}
\caption {Model performance and important features} \label{tab:R2} 
\begin{tabular}{| p{25pt} || p{25pt} |p{30pt} ||p{280pt} |} \hline
\textbf{Models} & \textbf{${R^2}$} & \textbf{Adj. ${R^2}$}  & \textbf{Important Features} (Ranked)\\\hline\hline
OLS & 0.09 & 0.08 & Security Policy, Pinned Dependencies, Maintained \\\hline
DT & 0.11 & 0.10 & Security Policy, Pinned Dependencies, Branch Protection, Maintained, Code Review \\\hline
GB & 0.12 & 0.10 & Code Review, Security Policy, Pinned Dependencies, Branch Protection, Maintained \\\hline
RF & -0.008 & -0.022 & Code Review, Pinned Dependencies, Branch Protection, Maintained, Security Policy \\\hline
\end{tabular}
\end{table*}
\subsubsection{\textbf{Model Performance:} }We measured the performance of each model with the following metrics:
 \begin{itemize}
     \item $\mathbf{R^2}$ is the statistical measure representing the proportion of the variance for a dependent variable explained by an independent variable or variables in a regression model. R-squared explains to what extent one variable's variance explains the second variable's variance or ``percent of variance explained by the model"~\cite{ratner2009correlation}. It supports values between 0 and 1, with greater values suggesting more variability explained by the model. A high $R^2$ value suggests good explanatory power but not predictive capability~\cite{zimmermann2008predicting}.
     \item The \textbf{adjusted} $\mathbf{ R^2}$ is a variant of R-squared that accounts for predictors that are not significant in a regression model. By considering the degrees of freedom of the independent variables and the sample population, adjusted $R^2$ demonstrates how well the data fit a curve or line and explains any bias in the $R^2$ value. A model's adjusted R-squared will decrease if insignificant variables are added and will increase when significant variables are added. Adjusted $R^2$ will always be less than or equal to $R^2$.
 \end{itemize}

\subsubsection{\textbf{Regression Model}} \label{RQ1_ans}
Four supervised machine learning models were used to evaluate the importance of each feature variable in interpreting the relationship between security practices and vulnerability count. Feature Importance techniques in machine learning models assign the score of input features for a given model based on their importance to predict the output. %

We first applied \textbf{ordinary least square (OLS)} to the dataset. Then, tree-based regression models were selected due to the non-normal distribution of the dataset. Since non-parametric tests do not require the data to follow the normal distribution~\cite{wei2007nonparametric}, we applied three tree-based regressor models in this study: \textbf{Decision tree (DT), Random Forest (RF)} and \textbf{Gradient Boosting (GB)}. Following a split of the dataset into a 70/30 training/test split, all models were fit on the training dataset. Then, we used grid search techniques to optimize each model's hyper-parameters. To minimize bias and over-fitting, ten-fold cross-validation was used during model training.

Table \ref{tab:R2} provides the performance comparison of four regression models. The dataset contains nine statistically significant variables suggested by the \texttt{SelectKBest} tool. %
Compared to all models, the Gradient Boosting (GB) model showed the highest ${R^2}$ and adjusted ${R^2}$ values, followed by DT, OLS and RF. However, the inferior performance of each model indicates that models (Table \ref{tab:R2}) were unable to explain the variability above 12\%. The OLS model found the highest coefficients for Security Policy, Pinned Dependencies, and Maintained. The DT model included Branch Protection and Code-Review as important features, along with Security Policy, Pinned Dependencies, and Maintained. The GB model ranked the same practices as DT as important practices. However, in the case of ranking, Code-Review was the most important metric in GB, whereas DT ranked Code-Review last. The RF model showed the lowest prediction performance of all four models, with a negative $R^2$ score. %

The performance of four models indicates that the currently-available independent security practices are inadequate to explain the variability of vulnerability counts. In Section \ref{challanges}, we discuss factors that might influence the model's prediction performance.

\subsection{\textbf{RQ2: Aggregate Security Score and Security Outcomes}} \label{RQ2_ans}
In this section, we describe our analysis for RQ2: \textit{\rqPattern} We look for the statistical association of a package's aggregate score with package vulnerability count to understand if implementing security practices assists in secure coding with fewer vulnerabilities. We hypothesize that- \textit{packages with higher security scores promote better security practices and contain fewer externally-reported vulnerabilities than packages with lower security scores}. We mentioned in Section~\ref{Scorecard_metrics} that Scorecard generates an aggregate score for each package, which is a weighted average of the individual checks, weighted by risk. To that end, we used the vulnerability count data of 2,422 packages (Section \ref{vulnerbilities}) as a target variable and the aggregate score of those packages as independent variables to answer RQ2. 

We removed six packages as outliers, identified by \texttt{qq-plot}. We perform the normality check to verify the normal distribution of the independent variable or aggregate score. We found the data was normally distributed to proceed with the linear regression model. The model, however, rejected our hypothesis by showing a significant positive relationship between aggregate score and vulnerability count. The  linear model (p-value $<$ 0.001, $R^2=0.04$, adj. $R^2=0.03$) had a 0.5-unit increase in vulnerability count for every unit increase in aggregate security score. 

\section{Discussion} \label{Discussion}
This section discusses the implications of our findings. We also addresses the other factors (section \ref{challanges}) that might impact the effectiveness of the use of Scorecard scores to identify the security practice. %
\subsection{\textbf{Important security practices}} \label{RQ1_disc}
To explore the findings related to RQ1, we used a wide range of security practices, multiple statistical learners, and prediction models to help practitioners to identify important security practices for improving package security. Table \ref{tab:R2} showed the performance of four models.%

Since we were interested in understanding the relationships between the variables, a low R-squared does not always negate the importance of any significant variables where there might be multiple unknown factors. Even with a low R-squared, statistically significant coefficients represent the mean change in the dependent variable given a one-unit shift in the independent variable~\cite{Low_R2,Low_R2_2}. Hence, using statistically significant variables as predictors indicates there is a relationship between these predictors and the vulnerability count in the current dataset. Our models explain the past data in a statistically significant manner, but its predictive ability is less than 12\% to extrapolate beyond the available data.

\textbf{Security Policy, Pinned Dependencies, Maintained, Code Review} and \textbf{Branch Protection} were the most important practices to explain our models. These five practices were statistically significant and came as an important feature in at least three models. Even if we discounted the RF model because of negative $R^2$, Security Policy, Pinned Dependencies, and Maintained came as an important practice in OLS, DT, GB models. Code Review and Branch Protection were important in DT and GB models. Our findings align with the Sonatype report~\cite{sonatype_8}, where they found \textbf{Code-Review, Pinned Dependencies, Branch Protection}, and \textbf{Binary-Artifacts} as important practices for Java projects. Binary-Artifacts were not important in our model for npm and PyPI ecosystems. Sonatype used 12k Java Projects, which is six times bigger than our dataset. We need more research to validate whether it is because of the size of the dataset or because Binary-Artifacts is a crucial practice for Java projects.

From correlation analysis~(Section ~\ref{cor}) we have observed that Pinned Dependencies is negatively co-related with four of these important practices (Security Policy, Maintained, Code Review, and Branch Protection). Pinned-Dependency prevents auto-updating a dependency to a new version without reviewing the differences between the two versions, which may include an insecure component. Zahan et al.~\cite{Zahan2022} showed that ``Pinned-Dependencies'' has high false positives in the Scorecard's current computing process, which might explain the inverse relationship. Scorecard looks for the presence of unpinned dependencies in Dockerfiles, shell scripts, and GitHub workflows. However, the tool does not check \texttt{package.json}, \texttt{package-lock.json} or \texttt{setup.py} files where npm and PyPI package developers lock their dependencies. Hence, ``Pinned-Dependencies'' scoring only involved GitHub-specific workflows, limiting the metric's generalizability for npm and PyPI ecosystems. To verify the negative association of ``Pinned-Dependencies" with other practices, we will need precise data from Scorecard tool that detects all pinned or unpinned dependencies in different ecosystems. Due to the necessity for additional research to establish the importance of ``Pinned-Dependencies" by gathering reliable data, we have thus withdrawn ``Pinned-Dependencies" as a suggested important practice. 

Scorecard defined Maintained, Code Review, and Branch Protection as high-risk practices and Security Policy as a medium-risk practice. Zahan et al.~\cite{Zahan2022} showed that both ecosystems have a gap in practicing Code-Review, Maintained, Branch-Protection, and Security-Policy practices. Although the current security trend shows practitioners do not employ these practices, our study indicates that practitioners adopting these practices will improve package security.

\begin{tcolorbox}[colback=black!10!white,colframe=black!10!white,boxrule=0.0mm, boxsep=0.1mm]
Maintained (high risk), Code Review (high risk), Branch Protection (high risk), and  Security Policy (medium risk) are the most important practices that practitioners can adopt to improve package security outcomes by minimizing vulnerabilities.
\end{tcolorbox}

\subsection{\textbf{Aggregate Security Score and Externally-reported Vulnerabilities:}}
Our model suggests that packages with more indicators of good security practices also had more reported vulnerabilities. Many factors are believed to increase the vulnerability count of packages. A possible explanation of the increase in reported vulnerability count while increasing in security score could be that the selected packages are used frequently by other open-source packages. Since more clients utilize these packages, there is a higher likelihood that the package will be tested or attacked, and there will be more reported vulnerabilities. The average stats (Direct Dependents: 427, Transitive Dependents: 2,758, Downloads: 72M) of packages in our dataset revealed that other open-source packages frequently used these packages. %
The number of developers may also impact the number of vulnerabilities. For instance, Meneely et al.~\cite{meneely2009secure} empirically showed that projects with more developers have more vulnerabilities. However, from OSV and Snyk databases, we can not collect such metadata to evaluate the relationship of package vulnerabilities with package popularity and developer community.  

Without exploring and controlling for confounding variables, such as package popularity, the findings are unlikely to provide certainty that the good security practices of these packages helped to find more vulnerabilities. We need to conduct a more thorough investigation to confirm the claim, taking into account the time it took to find and fix the vulnerabilities, the practices that were used to find them, and whether or not third-party users found the vulnerabilities while using the package rather than the package's owner. The scarcity of vulnerability data could be another reason influencing our model. We utilized externally-reported reported vulnerabilities and not those found prior to release and by the development team. %

\begin{tcolorbox}[colback=black!10!white,colframe=black!10!white,boxrule=0.0mm, boxsep=0.1mm]
Packages with increased good security practices also had increased reported vulnerability counts. Additional research is needed to determine if finding more vulnerabilities through good security practices is the cause of the increased vulnerability count or if the relationship is explained by other factors.
\end{tcolorbox}

\section{Challenges in available data } \label{challanges}
In this section, we discussed the challenges we found in explaining our findings. To improve model performance, future research replicating our study should consider these issues.
\subsection{\textbf{Scarcity of Vulnerability Data}} \label{scarcity_of_vulnerability}
Our findings raise the concern of scarce vulnerability data in OSS communities. We need more studies and accurate vulnerability data to evaluate the relationship between security practices and vulnerabilities as security outcomes, especially to comprehend how security practices influence vulnerability detection and prevention. A package may have more vulnerabilities than we found in OSV and Snyk vulnerability databases (Section ~\ref{vulnerbilities}). Dumidu et al.~\cite{wijayasekara2012mining} showed that a large number of vulnerabilities that affected the Linux kernel (39.4\%) and the MySQL database server (62.23\%) were vulnerabilities found in bug reports but not listed in common vulnerability databases, such as the NVD. Elder et al.~\cite{Elder2022} also identified 1,047 vulnerabilities in OpenMRS applications using five vulnerability detection techniques and tools. However, at the time of their study, OpenMRS had 11 reported CVE vulnerabilities. These studies show that vulnerability data that we acquire from OSV and Snyk databases hardly contain the accurate number of vulnerabilities as security outcomes. %

Additionally, even within our dataset, we removed four packages as outliers because the distance from other packages in terms of vulnerability count was high. Out of 2,272 packages, 2,268 had an average vulnerability count of 2.07; these four packages had an average of 122 vulnerabilities. Even if these four packages had real-world vulnerabilities that were more prevalent than those of other packages, we were unable to use the data to build models because of their low percentage (0.2\%) in the dataset. 

\subsection{\textbf{Security Score Data}}
Scorecards are designed to provide users of open-source projects with information that may be used to assess whether a dependency is secure. %
Although Scorecard collects raw/granular data to determine the normalized score, the public API only releases the security score of a package. A model using more granular data may have different outcomes from what we observed in this study. Studies show that defect prediction and vulnerability prediction can be improved using more granular data~\cite{madeyski2017continuous,jimenez2018engineering,zimmermann2008predicting}.

The Zahan et al.~\cite{Zahan2022} analysis also demonstrated that not all security practices scores accurately reflect the state of package security practices. For example, Dangerous-Workflow, Binary-Artifacts, Pinned-Dependencies, and Token-Permissions looked for the existence of specific properties in GitHub workflows. However, Scorecard does not verify the presence of GitHub workflows in a repository. Hence, if a package does not have GitHub workflows, the tool will grade that package as a score of 10 as part of good security practices. 

Then, the standards specified by the Scorecard tool for Dependency-Update-Tool, Fuzzing, and CII-Best-Practices were weakly adopted in npm and PyPI. One possible explanation behind weak adoption could be that these practices have inherent reliance on other system to measure package security. %
We cannot, however, avoid the possibility that packages may use different tools or techniques to improve security than the one defined by Scorecard. For example, 99\% of packages did not pass the CII-Best-Practices check because the maintainer did not use the CII Best Practices program to self-certify how they follow different best practices. %
Therefore, the lack of consensus on standardizing Scorecard metrics for different ecosystems is a challenge that Scorecard and the community should overcome to improve the security measures.

\subsection{\textbf{Time}}The multivariate nature of security practices significantly complicates its quantification and measurement~\cite{cheng2014metrics}. %
To validate security practice and security outcomes relationships, we need to evaluate the time when security practices were implemented and the time when vulnerabilities were injected, identified, and fixed. The data for time to remediate an externally-reported vulnerability can be collected from the vulnerabilities database. %
At the time of the study, Scorecard only measures security practices from the last 90 days' activities, which does not verify the security practices before 90 days. %
For better clarity, we need to do a controlled study by recording time of security practices implication in a packages and detection of vulnerabilities. The scorecard team run a weekly Scorecards scan and publish the results in a BigQuery public dataset. OSV and Snyk also record the time of vulnerability detection and fix. Researchers could gain valuable insight from our proposed model by conducting a long-term control study on package security practice data and tracking vulnerabilities.

\subsection{\textbf{Unpredictability}}Security metrics are a hard problem, especially in predicting vulnerabilities or assessing the effectiveness of counter measures~\cite{scala2019risk,cheng2014metrics}. We should consider that the software security field has an inherently greater amount of unexpected variation. Due to a lack of precision and unknowable variables, industries are still exploring how to adopt the most effective security measures to eliminate the most number of vulnerabilities efficiently~\cite{Elder2022,zhou2017automated,russell2018automated}. 

For instance, on July 20, 2022, Checkmarx security researchers warned people about a new supply chain attack strategy that uses fake commit metadata to give the appearance of legitimacy to malicious GitHub repositories~\cite{checkmarx}. Threat actors may modify the metadata in GitHub repositories to improve their reputation and increase the likelihood that application developers will choose them as reliable package owners. The researchers found it is possible to manipulate commit metadata, including timestamps, to make a repository appear older than it is or to show that credible contributors have been actively involved in its maintenance. This attack is an example of how frequently the software industry encounters new security threats. Attacks and attackers are unpredictable, as other recent supply chain attacks have also shown~\cite{Solarwinds_lesson,CODECOV}. Such unpredictability in software security often makes it challenging to interpret how security practices have affected software's security outcomes.

\section{Threats to Validity} \label{limitation}
To our knowledge, our study is the first peer-reviewed paper to look at how software security practices impact security outcomes using OpenSSF Scorecard data. However, as stated by Basili et al. ~\cite{basili1999building}, drawing general conclusions from empirical studies in software engineering is difficult. Because the software security process depends on a potentially large number of relevant context variables, we only used the OpenSSF Scorecard tool to measure software security practices. Our study findings are dependent on the ruleset defined by the tool. For this reason, we cannot assume a priori that the results of our study can be generalized to all ecosystems.

Since Scorecard only operated on GitHub repositories at the time of the study, GitHub was the only platform we looked at. Scorecard has four practices (Dangerous-Workflow, Binary-Artifacts, Pinned-Dependencies, and Token-Permissions) that are GitHub workflow specific. Hence, our findings could vary if we used dataset outside of GitHub repositories.

\section{Conclusion } In this study, we looked into how multiple security practices and their aggregate score affect the security outcome of software. First, our models identified Maintained, Code Review, Branch Protection and  Security Policy as the most important feature practices in our dataset. However, low $R^2$ and \texttt{adj} $R^2$ values raise concerns about model performance beyond the available data. Second, packages with increased good security practices also had increased reported vulnerability counts, demonstrating the need for confounding variable control and more vulnerability data to ascertain whether the relationship is explained by other factors. In the end, we explain our findings by discussing challenges related to scarce vulnerability data and security scores that might affect the model's performance. 

Future work should focus on collecting more granular-level data for security practices. The Scorecard team can contribute by releasing the non-normalized, granular data that they collect to determine the security score. Then, unsupervised machine learning can aid in identifying inherent trends in data to improve the model's performance. Additionally, we need more large-scale focus studies to identify accurate vulnerability data in software components for modeling and understand how security practices impact security outcomes. 

\section{ACKNOWLEDGMENTS}
This work was supported and funded by Cisco and National Science Foundation Grant No. 2207008. Any opinions expressed in this material are those of the author(s) and do not necessarily reflect the views of the National Science Foundation. We thank the OpenSSF Scorecard Team and Google GOSST for encouraging us to conduct this research. We also thank the NCSU Realsearch group for valuable feedback. In particular, we thank Parth Kanakiya for his assistance in generating vulnerability data.

\bibliographystyle{IEEEtran}
\bibliography{arxiv}

\begin{thebibliography}{10}
\providecommand{\url}[1]{#1}
\csname url@samestyle\endcsname
\providecommand{\newblock}{\relax}
\providecommand{\bibinfo}[2]{#2}
\providecommand{\BIBentrySTDinterwordspacing}{\spaceskip=0pt\relax}
\providecommand{\BIBentryALTinterwordstretchfactor}{4}
\providecommand{\BIBentryALTinterwordspacing}{\spaceskip=\fontdimen2\font plus
\BIBentryALTinterwordstretchfactor\fontdimen3\font minus
  \fontdimen4\font\relax}
\providecommand{\BIBforeignlanguage}[2]{{%
\expandafter\ifx\csname l@#1\endcsname\relax
\typeout{** WARNING: IEEEtran.bst: No hyphenation pattern has been}%
\typeout{** loaded for the language `#1'. Using the pattern for}%
\typeout{** the default language instead.}%
\else
\language=\csname l@#1\endcsname
\fi
#2}}
\providecommand{\BIBdecl}{\relax}
\BIBdecl

\bibitem{Synopsys_report}
\BIBentryALTinterwordspacing
Synopsys, ``2022 open source security and risk analysis,'' 2022. [Online].
  Available:
  \url{https://www.synopsys.com/software-integrity/resources/analyst-reports/open-source-security-risk-analysis.html}
\BIBentrySTDinterwordspacing

\bibitem{sonatype_2022}
Sonatype, ``700\% average increase in open source supply chain attacks,''
  \emph{https://www.sonatype.com/press-releases/sonatype-finds-700-average-increase-in-open-source-supply-chain-attacks},
  2022.

\bibitem{EO_2021}
D.~The White~House, Washington, ``Executive order on improving the nation’s
  cybersecurity,''
  \emph{https://www.federalregister.gov/documents/2021/05/17/2021-10460/improving-the-nations-cybersecurity},
  2021.

\bibitem{Solarwinds_lesson}
\BIBentryALTinterwordspacing
J.~M. Germain, ``Lessons learned from the solarwinds supply chain hack,'' 2021.
  [Online]. Available:
  \url{https://www.technewsworld.com/story/lessons-learned-from-the-solarwinds-supply-chain-hack-87029.html}
\BIBentrySTDinterwordspacing

\bibitem{CODECOV}
M.~JACKSON, ``Codecov supply chain breach - explained step by step,''
  \emph{https://blog.gitguardian.com/codecov-supply-chain-breach/}, 2021.

\bibitem{souppaya2022secure}
M.~Souppaya, K.~Scarfone, and D.~Dodson, ``Secure software development
  framework (ssdf) version 1.1,'' \emph{NIST Special Publication}, vol. 800, p.
  218, 2022.

\bibitem{NIST_2}
NIST, ``Software supply chain security guidance under executive order (eo)
  14028 section 4e,'' \emph{NIST Special Publication}, 2022.

\bibitem{Scorecard}
Scorecard, ``Security scorecards for open source projects,''
  \emph{https://github.com/ossf/scorecard}, 2021.

\bibitem{morrison2014mapping}
P.~Morrison, D.~Moye, and L.~A. Williams, ``Mapping the field of software
  security metrics,'' North Carolina State University. Dept. of Computer
  Science, Tech. Rep., 2014.

\bibitem{sec_outcome}
SANS, ``A sans 2021 survey: Rethinking the sec in devsecops: Security as
  code,''
  \emph{https://www.sans.org/webcasts/2021-survey-rethinking-sec-devsecops-security-code-118235/},
  2021.

\bibitem{morrison2015security}
P.~Morrison, ``A security practices evaluation framework,'' in \emph{2015
  IEEE/ACM 37th IEEE International Conference on Software Engineering},
  vol.~2.\hskip 1em plus 0.5em minus 0.4em\relax IEEE, 2015, pp. 935--938.

\bibitem{package_2}
J.~Crennan, ``What is a package registry?''
  \emph{https://blog.packagecloud.io/what-is-a-package-registry/}, 2021.

\bibitem{package_1}
S.~Jalan, ``An introduction to how javascript package managers work,''
  \emph{https://www.freecodecamp.org/news/javascript-package-managers-101-9afd926add0a/},
  2016.

\bibitem{scvs}
OWASP, ``Software component verification standard (scvs),''
  \emph{https://owasp-scvs.gitbook.io/scvs/}, 2020.

\bibitem{BSIMM}
BSIMM, ``Bsimm12 2021 foundations report,''
  \emph{https://www.bsimm.com/content/dam/bsimm/reports/bsimm12-foundations.pdf},
  2021.

\bibitem{weir2021infiltrating}
C.~Weir, S.~Migues, M.~Ware, and L.~Williams, ``Infiltrating security into
  development: exploring the world’s largest software security study,'' in
  \emph{Proceedings of the 29th ACM Joint Meeting on European Software
  Engineering Conference and Symposium on the Foundations of Software
  Engineering}, 2021, pp. 1326--1336.

\bibitem{CNCF_TAG}
C.~N.~C. Foundation, ``Software supply chain best practices,''
  \emph{https://github.com/cncf/tag-security/tree/main/supply-chain-security/supply-chain-security-paper},
  2021.

\bibitem{BSA}
BSA, ``The bsa framework for secure software: A new approach to securing the
  software lifecycle, version 1.1,''
  \emph{https://www.bsa.org/files/reports/bsa\_framework\_secure\_software\_update\_2020.pdf},
  2020.

\bibitem{OWASP_samm}
OWASP, ``Software assurance maturity model v2,''
  \emph{https://drive.google.com/file/d/1ZWMk4dpS3zpXjE28wi4cf5Lq6TUjeA5x},
  2020.

\bibitem{Zahan2022}
N.~Zahan \emph{et~al.}, ``Preprint: Can the openssf scorecard be used to
  measure the security posture of npm and pypi?'' \emph{arXiv preprint
  arXiv:2208.03412}, 2022.

\bibitem{SLSA}
SLSA, ``Supply chain levels for software artifacts,'' \emph{https://slsa.dev/},
  2022.

\bibitem{CISA_dev}
ESF, ``Securing the software supply chain: Recommended practices guide for
  developers,''
  \emph{https://www.cisa.gov/ESF\_SECURING\_THE\_SOFTWARE\_SUPPLY\_CHAIN\_DEVELOPERS.PDF},
  2022.

\bibitem{Microsoft_framework}
Microsoft, ``Oss ssc framework,''
  \emph{https://github.com/microsoft/oss-ssc-framework}, 2022.

\bibitem{npm}
OpenSSF, ``npm best practices guide,''
  \emph{https://github.com/ossf/package-manager-best-practices}, 2022.

\bibitem{OSI}
OSI, ``Open source insight(osi),'' \emph{https://deps.dev/}, 2022.

\bibitem{torres2019toto}
S.~Torres-Arias, H.~Afzali, T.~K. Kuppusamy, R.~Curtmola, and J.~Cappos,
  ``in-toto: Providing farm-to-table guarantees for bits and bytes,'' in
  \emph{28th USENIX Security Symposium (USENIX Security 19)}, 2019, pp.
  1393--1410.

\bibitem{in-toto}
in~toto, ``in-toto,'' \emph{https://in-toto.io/in-toto/}, 2021.

\bibitem{goggins2021open}
S.~Goggins, K.~Lumbard, and M.~Germonprez, ``Open source community health:
  Analytical metrics and their corresponding narratives,'' in \emph{2021
  IEEE/ACM 4th International Workshop on Software Health in Projects,
  Ecosystems and Communities (SoHeal)}.\hskip 1em plus 0.5em minus 0.4em\relax
  IEEE, 2021, pp. 25--33.

\bibitem{link2020open}
G.~J. Link, ``Open source project health,'' \emph{USENIX PATRONS}, p.~31, 2020.

\bibitem{sonatype_8}
Sonatype, ``2022 state of the software supply chain report,''
  \emph{https://www.sonatype.com/state-of-the-software-supply-chain/introduction},
  2022.

\bibitem{osv_adv}
Google, ``Open source vulnerability database,'' \emph{https://osv.dev/}, 2021.

\bibitem{Snyk_adv}
Snyk, ``Snyk vulnerability database,'' \emph{https://snyk.io/vuln}, 2021.

\bibitem{GHSA}
Github, ``Github advisory database,'' \emph{https://github.com/advisories},
  2021.

\bibitem{pypi_adv}
PyPA, ``Python packaging advisory database,''
  \emph{https://github.com/pypa/advisory-database}, 2021.

\bibitem{go_adv}
GoLang, ``The go vulnerability database,''
  \emph{https://github.com/golang/vulndb}, 2021.

\bibitem{rust_adv}
RustSec, ``Rustsec advisory database,''
  \emph{https://github.com/golang/vulndb}, 2018.

\bibitem{ossfuzz_adv}
Google, ``Oss-fuzz: Continuous fuzzing for open source software,''
  \emph{https://github.com/google/oss-fuzz}, 2016.

\bibitem{gsd_adv}
C.~S. Alliance, ``Global security database (gsd),''
  \emph{https://github.com/cloudsecurityalliance/gsd-database}, 2021.

\bibitem{zahan2022weak}
N.~Zahan, T.~Zimmermann, P.~Godefroid, B.~Murphy, C.~Maddila, and L.~Williams,
  ``What are weak links in the npm supply chain?'' in \emph{2022 IEEE/ACM 44th
  International Conference on Software Engineering: Software Engineering in
  Practice (ICSE-SEIP)}.\hskip 1em plus 0.5em minus 0.4em\relax IEEE, 2022, pp.
  331--340.

\bibitem{qq-plot}
Seaborn, ``seaborn-qqplot,''
  \emph{https://seaborn-qqplot.readthedocs.io/en/latest/}, 2022.

\bibitem{henze1990class}
N.~Henze \emph{et~al.}, ``A class of invariant consistent tests for
  multivariate normality,'' \emph{Communications in statistics-Theory and
  Methods}, vol.~19, no.~10, pp. 3595--3617, 1990.

\bibitem{zar2005spearman}
J.~H. Zar, ``Spearman rank correlation,'' \emph{Encyclopedia of biostatistics},
  vol.~7, 2005.

\bibitem{ratner2009correlation}
B.~Ratner, ``The correlation coefficient: Its values range between+ 1/- 1, or
  do they?'' \emph{Journal of targeting, measurement and analysis for
  marketing}, vol.~17, no.~2, pp. 139--142, 2009.

\bibitem{selectK}
sklearn, ``Selectkbest,''
  \emph{https://scikit-learn.org/stable/modules/generated/sklearn.feature\_selection.SelectKBest.html},
  2022.

\bibitem{zimmermann2008predicting}
T.~Zimmermann \emph{et~al.}, ``Predicting defects using network analysis on
  dependency graphs,'' in \emph{Proceedings of the 30th international
  conference on Software engineering}, 2008, pp. 531--540.

\bibitem{wei2007nonparametric}
Z.~Wei \emph{et~al.}, ``Nonparametric pathway-based regression models for
  analysis of genomic data,'' \emph{Biostatistics}, vol.~8, no.~2, pp.
  265--284, 2007.

\bibitem{Low_R2}
J.~Frost, ``How to interpret regression models that have significant variables
  but a low r-squared,''
  \emph{https://statisticsbyjim.com/regression/low-r-squared-regression/},
  2022.

\bibitem{Low_R2_2}
K.~Grace-Martin, ``Can a regression model with a small r-squared be useful?''
  \emph{https://www.theanalysisfactor.com/small-r-squared/}, 2022.

\bibitem{meneely2009secure}
A.~Meneely and L.~Williams, ``Secure open source collaboration: an empirical
  study of linus' law,'' in \emph{Proceedings of the 16th ACM conference on
  Computer and communications security}, 2009, pp. 453--462.

\bibitem{wijayasekara2012mining}
D.~Wijayasekara, M.~Manic, J.~L. Wright, and M.~McQueen, ``Mining bug databases
  for unidentified software vulnerabilities,'' in \emph{2012 5th International
  conference on human system interactions}.\hskip 1em plus 0.5em minus
  0.4em\relax IEEE, 2012, pp. 89--96.

\bibitem{Elder2022}
S.~Elder \emph{et~al.}, ``Do i really need all this work to find
  vulnerabilities? an empirical case study comparing vulnerability detection
  techniques on a java application,'' \emph{arXiv preprint arXiv:2208.01595},
  2022.

\bibitem{madeyski2017continuous}
L.~Madeyski and M.~Kawalerowicz, ``Continuous defect prediction: the idea and a
  related dataset,'' in \emph{2017 IEEE/ACM 14th International Conference on
  Mining Software Repositories (MSR)}.\hskip 1em plus 0.5em minus 0.4em\relax
  IEEE, 2017, pp. 515--518.

\bibitem{jimenez2018engineering}
M.~Jimenez \emph{et~al.}, ``[engineering paper] enabling the continuous
  analysis of security vulnerabilities with vuldata7,'' in \emph{2018 IEEE 18th
  International Working Conference on Source Code Analysis and Manipulation
  (SCAM)}.\hskip 1em plus 0.5em minus 0.4em\relax IEEE, 2018, pp. 56--61.

\bibitem{cheng2014metrics}
Y.~Cheng, J.~Deng, J.~Li, S.~A. DeLoach, A.~Singhal, and X.~Ou, ``Metrics of
  security,'' in \emph{Cyber defense and situational awareness}.\hskip 1em plus
  0.5em minus 0.4em\relax Springer, 2014, pp. 263--295.

\bibitem{scala2019risk}
N.~M. Scala, A.~C. Reilly, P.~L. Goethals, and M.~Cukier, ``Risk and the five
  hard problems of cybersecurity,'' \emph{Risk Analysis}, vol.~39, no.~10, pp.
  2119--2126, 2019.

\bibitem{zhou2017automated}
Y.~Zhou and A.~Sharma, ``Automated identification of security issues from
  commit messages and bug reports,'' in \emph{Proceedings of the 2017 11th
  joint meeting on foundations of software engineering}, 2017, pp. 914--919.

\bibitem{russell2018automated}
R.~Russell, L.~Kim, L.~Hamilton, T.~Lazovich, J.~Harer, O.~Ozdemir,
  P.~Ellingwood, and M.~McConley, ``Automated vulnerability detection in source
  code using deep representation learning,'' in \emph{2018 17th IEEE
  international conference on machine learning and applications (ICMLA)}.\hskip
  1em plus 0.5em minus 0.4em\relax IEEE, 2018, pp. 757--762.

\bibitem{checkmarx}
I.~Arghire, ``Supply chain attack technique spoofs github commit metadata,''
  \emph{https://www.securityweek.com/supply-chain-attack-technique-spoofs-github-commit-metadata},
  2022.

\bibitem{basili1999building}
V.~R. Basili \emph{et~al.}, ``Building knowledge through families of
  experiments,'' \emph{IEEE Transactions on Software Engineering}, vol.~25,
  no.~4, pp. 456--473, 1999.

\end{thebibliography}
\end{document}